\documentclass[10pt,letterpaper]{article}
\usepackage[top=0.85in,left=2.75in,footskip=0.75in,marginparwidth=2in]{geometry}

\usepackage[utf8]{inputenc}

\usepackage{cite}

\usepackage{nameref,hyperref}

\usepackage{microtype}
\DisableLigatures[f]{encoding = *, family = * }

\raggedright
\usepackage{changepage}

\usepackage[aboveskip=1pt,labelfont=bf,labelsep=period,singlelinecheck=off]{caption}

\makeatletter
\renewcommand{\@biblabel}[1]{\quad#1.}
\makeatother

\usepackage{lastpage,fancyhdr,graphicx}
\usepackage{epstopdf}
\fancyheadoffset[L]{2.25in}
\fancyfootoffset[L]{2.25in}

\usepackage{color}

\definecolor{Gray}{gray}{.25}

\usepackage{graphicx}

\usepackage{sidecap}

\usepackage{wrapfig}
\usepackage[pscoord]{eso-pic}
\usepackage[fulladjust]{marginnote}
\reversemarginpar

\begin{document}
\vspace*{0.35in}

\begin{flushleft}
{\Large
\textbf\newline{On the physical mechanisms underlying single molecule dynamics in simple liquids}
}
\newline
\\
R. G. Keanini\textsuperscript{1,*},
Jerry Dahlberg\textsuperscript{1},
Peter T. Tkacik\textsuperscript{1}
\\
\bigskip
\bf{1} Department of Mechanical Engineering,  University of North Carolina at Charlotte, Charlotte, NC  28223
\\
\bigskip
* rkeanini@uncc.edu

\end{flushleft}

\section*{Abstract}
Physical arguments and comparisons with published experimental data suggest that in simple liquids: i) single-molecule-scale viscous forces are produced by temperature-dependent London dispersion forces, ii) viscosity decay with increasing temperature reflects electron cloud compression and attendant suppression of electron screening, produced by increased nuclear agitation, and iii) temperature-dependent self-diffusion is driven by a narrow band of phonon frequencies lying at the low-frequency end of the solid-state-like phonon spectrum. The results suggest that collision-induced electron cloud distortion plays a decisive role in single molecule dynamics: i) electron cloud compression produces short-lived repulsive states and single molecule, self-diffusive hops, while ii) shear-induced distortion generates viscosity and single-molecule-scale viscous drag. The results provide new insight into nonequilibrium molecular dynamics in nonpolar, nonmetallic liquids.

\section*{Introduction}
Gaining a deeper understanding of single molecule dynamics in liquids bears on fundamental problems in chemical reaction kinetics \cite{chemrxn1, chemrxn2}, sub-cellular water transport \cite{waterxport1, waterxport2} and biomass transfer \cite{biomass1, biomass2}, detection of cosmic particles and radiation \cite{cosmic1}, dark matter detection \cite{darkmatter1, darkmatter2, darkmatter3}, detection of collision products in high energy physics \cite{cosmic1}, corrosion \cite{corrosion}, and weathering of terrestrial and extra-terrestrial surface rock \cite{rock}.  The problem has attracted the attention of luminaries like Einstein \cite{einstein1, einstein2}, Perrin \cite{perrin}, Laundau \cite{landau}, Prigogine \cite{prigogine}, and Feynman \cite{feynman}. Nonetheless, the physical mechanisms that determine single molecule motion in liquids remain poorly understood.

A variety of experimental and theoretical approaches have been developed for studying molecular dynamics in liquids. Experimental techniques include light and particle scattering \cite{lovesey1, lovesey2, bernepecora, boonyip, skoldrowe} which probes dynamic responses over single- to multiple-molecule length-scales, and sub-collision and longer $ \left( t \geq \mathrm{O} \left( 10^{-14} \ \mathrm{s} \right) \right) $ time scales. Photonic techniques \cite{photonics1, photonics2, photonics3, photonics4, photonics5} are capable of exposing intramolecular dynamics on femtosecond time scales $ \left( \mathrm{t} = \mathrm{O} \left( 10^{-15} \ \mathrm{s} \right) \right) . $ Molecular dynamics simulations provide a computational approach for probing each of these scales \cite{rahman, levesque, evansmorriss}.

Theoretical modeling drives and allows interpretation of typically complicated experimental observations. Over multi-molecule length-scales and multiple-collision time scales, molecular hydrodynamics \cite{lovesey2, bernepecora, boonyip, mountain, forster, hansen, evansmorriss} successfully connects observed spectral responses of dense fluids to the continuum Navier-Stokes (NS) equations. However, on single-molecule length scales and collision- and sub-collision time scales, mapping molecular-scale response into generalized NS models requires time- and space-dependent transport coefficients \cite{boonyip, forster, hansen, lovesey2, kubo}, revealing our poor understanding of single-molecule liquid-state dynamics. 

For atomic, and small polyatomic, nominally spherical, nonpolar liquids, Langevin (LE) models \cite{forster, boonyip, hansen} provide a powerful, particle-based framework for tackling molecular dynamics problems, both under classical conditions - where the dynamical processes of interest take place on time scales exceeding the 'dispersion time scale', $ \tau_d = O \left( 10^{-16} \ \mathrm{s} \right) ,$ see below - and under conditions where quantum smearing of the dynamics becomes important \cite{zwanzig, forster}.  Generally speaking, LE models are suitable for particle dynamics problems characterized by short-time scale random forcing and longer-time scale non-random dynamical dissipation, as well as by possible external forcing.

This paper presents three results, which together, provide new insight into the dynamics of single molecules in nonpolar liquids:

\vspace{0.2cm}

\noindent a) A simple physical model is proposed which suggests that: i) liquid-state viscosity is produced by temperature-dependent London dispersion forces, and ii) viscosity decay with increasing temperature reflects decreased  electron screening of nuclear charge.  Comparison of predicted and experimentally observed viscosities \cite{rouschviscosity, boonyip} for liquid Ne, Ar, Kr, Xe, $ \mathrm{N_2},$ $ \mathrm{O_2} ,$ and $ \mathrm{CH_4} ,$ support the proposed physical picture. 

\vspace{0.2cm}

\noindent b) A Langevin model of sub-collision time scale, single molecule dynamics, which explicitly accounts for solid-state-like phonon modes, leads to a physically consistent explanation for self- diffusion coefficients measured in liquid Ar, Kr, and Xe \cite{kinetictheoryX}. The model suggests: i) that on time scales ranging from the Frenkel scale, $ \tau_F = O \left( 10^{-14} \ \mathrm{s} \right)  $ - where $ \tau_F $ is approximately an order of magnitude shorter than the characteristic intermolecular collision time scale, $ \tau_c = O \left( 10^{-13}  \ \mathrm{s} \right) - $ down to the fast dispersion scale, $ \tau_d = O \left( 10^{-16} \ \mathrm{s} \right)  , $ molecular dynamics in simple liquids is solid-like, and thus dominated by phonon modes, consistent with the equilibrium statistical mechanics picture presented by \cite{bolmatov, phononliqreview}, and ii) that the random diffusional hopping of individual molecules is produced by a narrow band of phonon modes lying near the low-frequency end of the phonon spectrum, $ \omega_c \sim \omega_F = 2 \pi / \tau_F .$

\vspace{0.2cm}

\noindent c) A set of time scale-dependent Langevin equations are proposed for describing single molecule dynamics in simple, non-polar liquids.  The equations apply over the poorly characterized sub-collision time scale, $ \tau_d \leq t \leq \tau_c , $ incorporate the above results, and represent best-guess extrapolations of well-established dynamics on longer time scales. 

As a consequence of the modeling, experimental comparisons, and consistency checks that are presented, we arrive at a preliminary picture of the decisive role apparently played by collision-induced electron cloud distortion in single molecule dynamics. Arguments and evidence are presented suggesting that phonon-induced electron cloud compression can force colliding molecular pairs into short-lived repulsive states, producing, in turn, single molecule, self-diffusive hops. In addition, we propose that nonequilibrium, shear-induced, 'tangential' electron cloud distortions generate viscosity and single molecule scale, resistive viscous forces.    

Aside 1: The term 'molecular' will refer to monatomic as well as small, polyatomic liquids.  \\
\noindent Aside 2: In order to provide a physical feel for the important time scales in this problem, we will often use those associated with liquid Ar.

\section*{Results, Methods and Discussion}
\section*{Dispersion forces and electron screening determine temperature-dependent dynamic viscosity}
As a measurable property determined by molecular-scale processes, viscosity provides a window into molecular dynamics. Here, we study temperature-dependent viscosities observed in liquid Ne, Ar, Kr, Xe, $ \mathrm{N_2} ,$ $ \mathrm{O_2} ,$ and $ \mathrm{CH_4 }, $ at fixed pressures, over the temperature ranges on which each specie exists as a liquid \cite{boonyip, rouschviscosity}.  

The corresponding states principle \cite{hirschfelder, boonyip, improvedcorrstates} provides the basis for our argument. The simplest form of the principle postulates that viscosity is determined by a characteristic intermolecular potential energy, $ \epsilon , $ a characteristic intermolecular length-scale, $ \sigma , $ the molecular mass, $ M , $ and a specie-dependent temperature-scale, $ \epsilon / k_B : $ 
\begin{equation}\label{mucorrstateeqn}
\mu = f \left( T , M \epsilon, \sigma , k_B \right) 
\end{equation}
where, on dimensional grounds, $ M $ can be grouped with $ \epsilon . $ 
Dimensional analysis allows restatement of (\ref{mucorrstateeqn}) in  nondimensional form: \begin{equation}\label{nondimmueqn}
\tilde{\mu} = \tilde{f} \left( \tilde{T} \right) 
\end{equation}
where $ \tilde{\mu} = \mu / \sqrt{M \epsilon} / \sigma^{2} , $ $ \tilde{T} = T / \left(\epsilon / k_B \right) , $ $ k_B $ is Boltzmann's constant, and $ \tilde{f} $ represents the experimentally determined correlation. The principle holds nominally well \cite{boonyip, hirschfelder} in nonpolar atomic and diatomic liquids that are well-modeled by the Lennard-Jones potential \cite{hirschfelder, boonyip}. More comprehensive correlations incorporating quantum (low temperature and small mass) effects and information on the shape of the intermolecular potential have been proposed \cite{improvedcorrstates}. 

In order to derive what turns out to be a simple physical model for predicting viscosity in nonpolar liquids, we proceed in three steps. First, a scaling argument is used to place the corresponding states principle on a physical basis, leading to an approximate relationship for $ \mu:  $
\begin{equation}\label{viscosityscale}
\mu \approx \frac{\sqrt{ \epsilon M}}{\sigma^2}
\end{equation}
Focusing on nonpolar liquids subject to London dispersion forces, we then follow \cite{hirschfelder} and state the intermolecular energy, $ \epsilon ,$ in terms of specie polarizability, $ \alpha ,$ and the separation, $ r_{ab} , $ between colliding molecular pairs. Finally, collision-induced, and temperature-dependent polarization is stated in terms a mean, temperature-dependent electron cloud distortion, $ \delta \sigma \left( T \right) . $ Importantly, in order to obtain a viscosity model consistent with available measurements \cite{boonyip}, we propose that electron cloud distortion, $ \delta \sigma \left( T \right) ,$ decreases linearly with temperature.

Confine attention to classical conditions, assume pair-wise intermolecular collisions - see point i) in the final section below, neglect non-spherical shape effects \cite{improvedcorrstates} on the (pair-wise) intermolecular potential, and focus on simple, nonmetallic liquids, i.e., those composed of nonpolar molecules having nominally spherical, localized electron distributions \cite{ingebritsen}.  Under these conditions, the \textit{attractive} potential between colliding molecular pairs is wholly determined by London dispersion
\cite{london, hirschfelder}. 

On time scales longer than the dispersion time scale, $ \tau_d = \mathrm{O} \left( 10^{-16} \ \mathrm{s} \right) $ - the scale on which electron distributions oscillate
\cite{french} - but shorter than the intermolecular collision time scale, $ \tau_c = \mathrm{O} \left( 10^{-13} \ \mathrm{s} \right), $ two-body interactions dominate three- and higher-order interactions; again, see point i) in the final section. Due to  high-frequency phonon modes, $ \omega > 2 \pi \tau_F^{-1} = \mathrm{O} \left( 10^{14} \ \mathrm{s^{-1}}  \right), $ we assume that viscosity emerges on an intermediate time scale, $ \tau_v , $ where $ \tau_d << \tau_v << \tau_c . $ 
As shown by \cite{volumestokeseinstein}, a modified Stokes-Einstein relation,
\begin{equation}\label{stokeseinstein}
D = \frac{2 k_B T}{ n' \pi \mu \sigma f' }
\end{equation}
connects diffusion of small and medium sized molecules (in water and carbon tetrachloride, \cite{volumestokeseinstein}) to the viscosity of the solvent liquids. Here, $ D $ is the diffusion coefficient, $ \mu$ is the dynamic viscosity, $ f' $ is a molecular-shape-dependent factor, and $ n' $ is a correction factor, ranging from approximately 2 to 6, and accounting for a mix of slip- and no-slip flow conditions on a molecule's surface.

Importantly, (\ref{stokeseinstein}) implies that single molecule dynamics can be modeled using the simple, memory free Langevin equation:
\begin{equation}\label{langevineqnsimple}
M \frac{ d \mathbf{v} \left( t \right)}{dt} = - 3 \pi \sigma_m \mu \mathbf{v} + \mathbf{F_R} \left( t \right) 
\end{equation}
where $ \sigma_m = n'f' \sigma / 6 $ is the effective molecular diameter;
see, e.g., \cite{chandrasekhar}. Here, $ \mathbf{v} \left( t \right) $ is the instantaneous velocity of the molecule and $ \mathbf{F_R} \left(  t \right) $ is the instantaneous random force on the molecule. Thus, on time scales of order $ \tau_v = \mathrm{O} \left( 10^{-15} \ \mathrm{s} \right) , $ and longer, we argue that the rate of work done on an individual molecule by the dispersion force is dissipated by viscous dissipation:
\begin{equation}\label{dispersionandviscousforce}
\frac{\epsilon}{\sigma} \cdot \delta r_{nuc} \cdot \tau_v^{-1} \approx \mu \frac{u_{nuc}}{\sigma} \cdot \sigma^2 \cdot \delta r_{nuc} \cdot \tau_v^{-1}
\end{equation}
where spatial derivatives are approximated as $ 1 / \sigma , $ the characteristic speed of the nucleus is given by $ u_{nuc} \approx \sqrt{\epsilon / M }, $ the characteristic nuclear displacement over $ \tau_v $ is represented as $ \delta r_{nuc} , $ and the nominal surface area of the molecule is on the order of $ \sigma^2 . $ Solving (\ref{dispersionandviscousforce}) for the viscosity then leads to (\ref{viscosityscale}).

\subsection*{Dispersion forces determine viscosity and increased nuclear agitation with temperature compress electron clouds, suppressing viscosity}
In simple liquids subject to London interactions, the energy scale, $ \epsilon , $ is approximately determined by \cite{london, hirschfelder}
\begin{equation}\label{dispeqn} 
\epsilon_d = \frac{3}{4} h \nu_o \frac{\alpha^2}{r_{ab}^6}
\end{equation}
where $ h \nu_o $ is the ground state energy  of an isotropic quantum oscillator, $ \alpha $ is the polarizability, and $ r_{ab} $ is the separation between the molecular pair's nuclei. This expression follows from
assuming that pair-wise molecular collisions correspond to weak interactions between isotropic quantum oscillators \cite{london, hirschfelder}.  As an initial  consistency check, Appendix 1 compares estimated and experimental kinematic viscosities, $ \nu = \mu / \rho , $ for a set of simple liquids, where $\nu$ estimates use
London's rigorous second order quantum perturbation model \cite{london, hirschfelder}, $ \epsilon = C / r_{\mathcal{A} \mathcal{B}}^6 $ in (\ref{viscosityscale}), and where $ C $ is the attractive constant.  

Since (\ref{dispeqn}) allows intuitive derivation and interpretation of the viscosity estimate presented here,
as well as exposing the apparent central role of electron cloud distortion in viscosity generation, we use (\ref{dispeqn}) to estimate $ \epsilon . $ Polarizability is given approximately by \cite{hirschfelder}
\begin{equation}\label{polareqn}
\alpha = \frac{4}{9 a_o } \sum_{i=1}^n \left( \overline{r_i^2} \right)^2
\end{equation}
where the sum is taken over the principle quantum energy levels of a given molecule, $ \overline{r_i^2} $ is the average squared displacement of the electrons in $ i^{th} $ shell (induced by an external electric field), and $a_o $ is the Bohr radius. In detail, $ \overline{r_i^2}, $ follows from introduction of Slater orbitals \cite{slater, hirschfelder}:
\begin{equation}\label{rsquaredformula}
\overline{r_i^2}=\left[ \frac{n_i^* }{ 2\left( Z - S_i \right) } \right]^2 \left(2 n_i^* + 1 \right) \left(2 n_i^* + 2 \right) a_o^2
\end{equation}
where $ Z - S_i $ is the effective nuclear charge of the $ i^th $ shell, $ Z $ is the nuclear charge, $ S_i $ is the associated screening constant, and $ n_i^* $ the effective principle quantum number.

Focusing on Ne, Ar, Kr, Xe, $ \mathrm{O_2} , $ and $ \mathrm{N_2}, $ 
we label the sum of mean squared electron displacements as
\begin{equation}\label{rsqureddefn}
\frac{1}{n} \sum_{i=1}^n \left( \overline{r_i^2} \right)^2 = \delta \sigma^4 \left( T \right)
\end{equation}
where 
\begin{equation}\label{meanelectrondisplacement}
\delta \sigma \left( T \right) = \left[ n^{-1} \sum_{i=1}^n \left( \overline{r_i^2} \right)^2 \right]^{1/4}
\end{equation}
represents the average collision-induced distortion of all the electrons in a molecule, and where we assume that $ \delta \sigma \left( T \right) $ is temperature-dependent. 

Over the narrow temperature ranges on which each of these species exist as a liquid, and based on the observation that liquid viscosities decrease with increasing temperature \cite{boonyip}, we introduce an ansatz that the mean electron distortion decreases linearly with temperature:
\begin{equation}\label{polartemp}
\delta \sigma \left( T \right) = \delta \sigma_v 
\left( 1 - \epsilon_T \right) = \delta \sigma_v \left( 1 - \frac{T_v-T}{T_v} \right)
\end{equation}
where $ \delta \sigma_v = \delta \sigma \left( T_v \right) , $ is the mean displacement at characteristic temperature, $ T_v = \epsilon / k_B , $ and $ \epsilon_T = \left(T_v - T \right) / T_v .$  

Physically, and in light of (\ref{rsquaredformula}) and the results below, this guess suggests that electron screening \textit{decreases} with increasing temperature, consistent with
behavior observed in deuterated metals \cite{electronscreening}. Since kinetic energy of both nuclei and electrons increase with rising temperature, where the latter presumably enhances screening, the suppression of screening apparently reflects increased nuclear agitation; intensifying agitation, under spatially packed conditions effectively thins surrounding electron clouds.  This mechanism may also underlie atomization of vapor phase molecular clusters as $ T \rightarrow T_v = \epsilon / k_B ,$ where atomization reflects nuclear kinetic energy overtaking intermolecular dispersion forces.   

Next, approximate $ \delta \sigma \left( T \right) = \delta \sigma_v \left( 1 - \epsilon_T \right) $ as $ \delta \sigma \left( T \right) = \delta \sigma_v \exp \left( - \epsilon_T \right) = \delta \sigma_v e^{1} \exp \left( -T^* \right) , $ where $ T^* = T / T_v = T / \left( \epsilon / k_B \right) , $ is the dimensionless temperature defined in the corresponding states correlation. Since the maximum magnitude of $ \epsilon_T $ is on the order of 0.3 for the set of liquids considered, save oxygen, the maximum error introduced by replacing $ \left( 1 - \epsilon_T \right) $ with $ \exp \left( - \epsilon_T \right) $ is on the order of 10 \%.  

Using $ \epsilon_d $ in (\ref{dispeqn}) for $ \epsilon $ in (\ref{viscosityscale}), the definition in (\ref{meanelectrondisplacement}) for the mean electron distortion, and the exponential approximation above for the assumed linear temperature variation in $ \delta \sigma \left( T \right) ,$ leads to an approximate expression for the temperature-dependent viscosity for simple liquids:
\begin{equation}\label{viscosityfinal}
\mu \left( T^* \right) \approx C_o \exp \left( -4 T^* \right) 
\end{equation}
where $ C_o = \sqrt{ 243 h \nu_o e^8 \left( \delta \sigma_v / \sigma_o \right)^8 M /16 } / \sigma_o^2 ,$ and where $ a_o $ is approximated as $ \sigma_o /2 ,$ and $ \sigma_o $ corresponds, e.g., to the molecular diameter at the specie melting point. 

\subsection*{Comparison of theoretical and observed viscosities}
In order to allow comparison of theoretical, temperature-dependent viscosities, as given by (\ref{viscosityfinal}), with experimentally measured viscosities, we define the dimensionless viscosity for liquid $ \kappa , $ $ \mu_{\kappa}^* \left( T^* \right) = \mu_{\kappa} \left( T^* \right) / \mu_{m, \kappa} ,$ yielding
\begin{equation}\label{dimlessnu}
\mu_{\alpha}^* \left( T^* \right) = c_{\kappa}^* \exp \left( -4 T^* \right) 
\end{equation}
where $ c_{\kappa}^* = \mu_{\kappa}^* \left( T_{min, \kappa}^* \right) \exp \left( 4 T_{min, \kappa}^* \right) , $ and where $ \mu_{\kappa}^* \left( T_{min, \kappa}^* \right) $ is the measured dimensionless viscosity for fluid $  \kappa $ at the minimum dimensionless temperature, $ T_{min, \kappa}^* $ at which $ \mu_{\kappa}^* $ is measured.  Note, due to the approximations used to obtain the constant $ C_o $
above, plotted viscosity estimates obtained using (\ref{viscosityfinal}) exhibit the appropriate decay with temperature, but are displaced by a (nominally) fixed magnitude from measured viscosities.

\begin{figure}[!tbhp]
  \centering
    \includegraphics[width=\textwidth, trim = {1.5cm 1cm 1.5cm 2.5cm}]{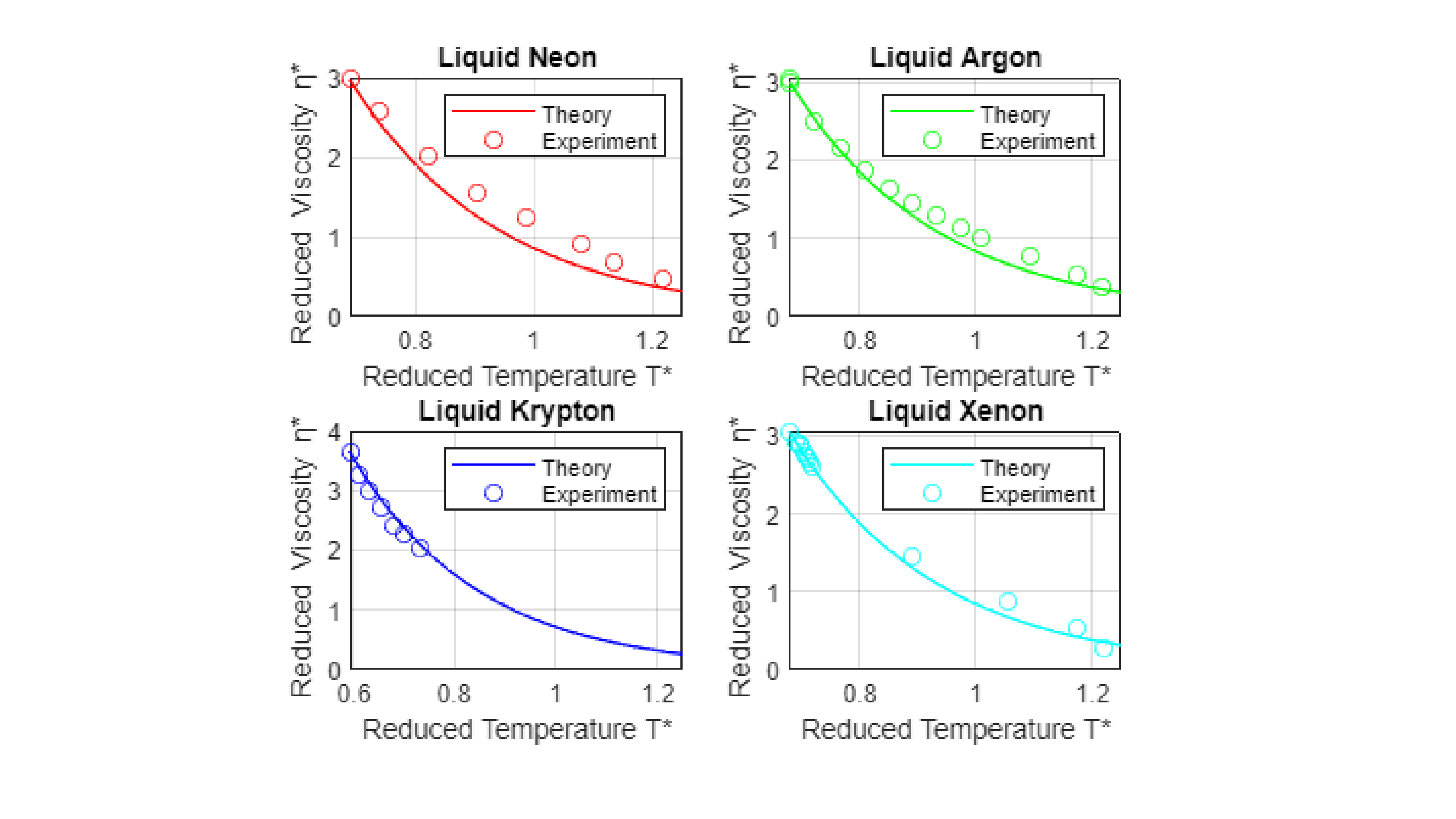}
    \caption{Temperature-dependent viscosity for noble liquids. The proposed model assumes: i) dominant pairwise, dispersive, intermolecular interactions - see the scaling argument \textbf{i)} in the final section, and ii) that the average collision-induced distortion of the molecule's electrons, $ \delta \sigma \left( T \right) = \left[ \sum_{i=1}^n \left( \overline{r_i^2} \right)^2 \right]^{1/4},$  decays linearly with temperature. The second assumption suggests that electron screening \textit{decreases} with increasing temperature - consistent with \cite{electronscreening} - and, in turn, that thermally-driven nuclear motion dominates presumed enhanced electron shrouding of the nucleus.  For an explanation of experimental conditions and definitions of dimensionless variables, see the caption to Fig. \ref{n2o2ch4viscosity}.}
 \label{nobleviscosity}
\end{figure}

\begin{figure}[!tbhp]
  \centering
     \includegraphics[width=\textwidth, trim = {1.5cm 1cm 1.5cm 1.5cm}]{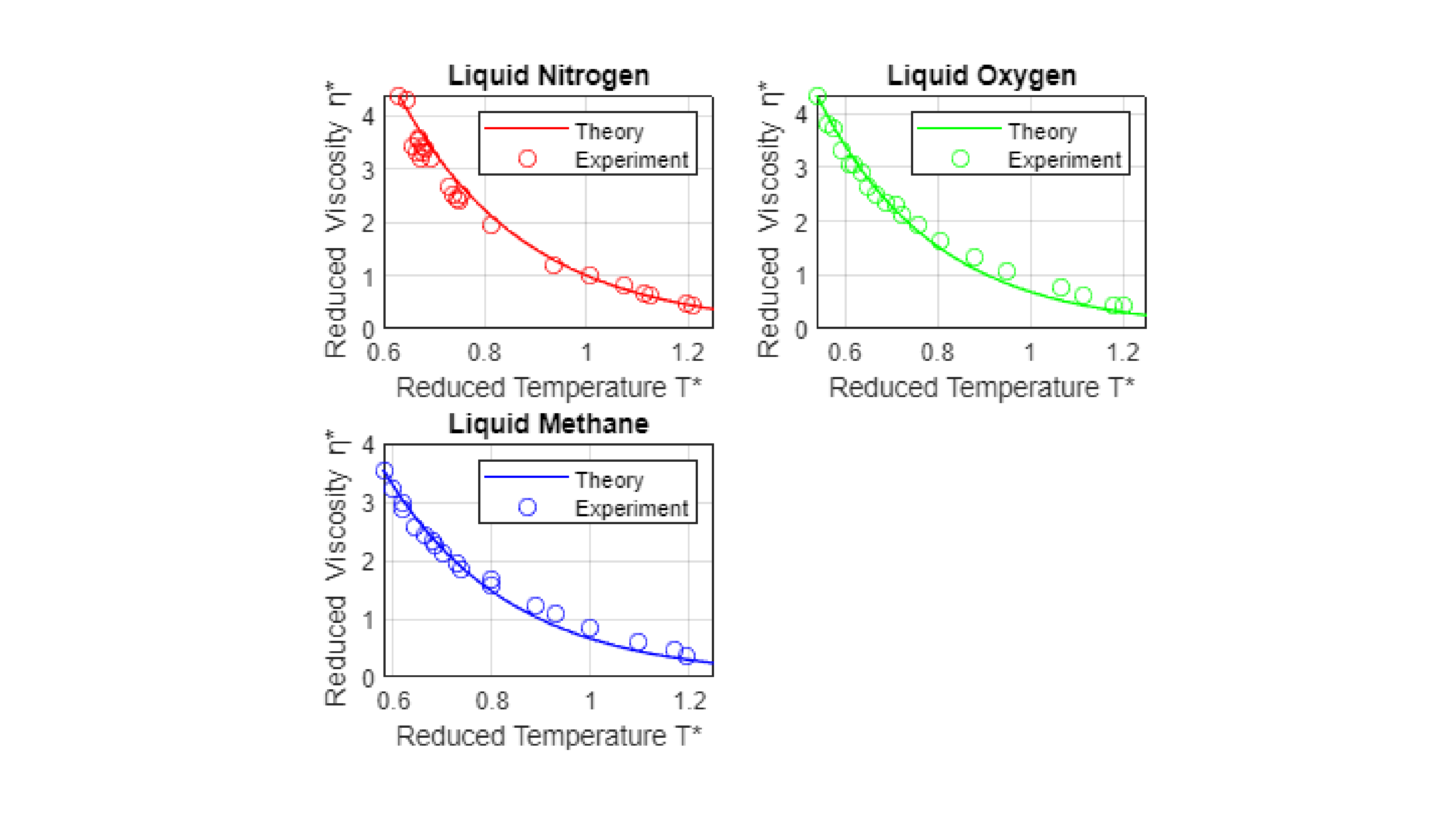}
     \caption{Temperature-dependent liquid viscosities for liquid $ \mathrm{N_2} ,$ $ \mathrm{O_2} ,$ and $ \mathrm{CH_4} .$ Notes: a) Experimental viscosity for specie $ \kappa $ is nondimensionalized using the viscosity scale $ \mu_{scale, \kappa} = \sqrt{M_{\kappa} \epsilon_{min, \kappa} } \sigma_{min, \kappa}^2 , $ where  $ \epsilon_{min, \kappa} , $ is the minimum Lennard-Jones potential,  $ \sigma_{min, \kappa} , $ is the molecular separation at which the potential is minimized, and  $ M_{\kappa} $ is the molecular mass.  Likewise, nondimensional experimental temperatures are scaled using $ T_{scale,\kappa} = \epsilon_{min, \kappa} / k_B , $ where $ k_B $ is Boltzmann's constant.  b) The minimum temperature, $ T_{min, \kappa} , $ at which each specie viscosity was measured corresponds approximately to the melting temperature (at atmospheric pressure) for that specie. c) The maximum measurement temperature for each specie, $ T_{max, \kappa} , $ in all six cases, exceeds the specie's atmospheric boiling point. Thus, the viscosities shown exceed the range of temperatures over which each specie is in the liquid state. d)  References describing the experimental techniques used in measuring temperature-dependent viscosities can be found in \cite{rouschviscosity}.}
    \label{n2o2ch4viscosity}
\end{figure}

Theoretical and experimental, temperature dependent dimensionless viscosities for the six simple liquids considered here are compared in Figs. \ref{nobleviscosity} and \ref{n2o2ch4viscosity}.  The comparisons lend significant support to our central argument: Decaying liquid viscosity in simple liquids reflects decreased electron screening of the positive nucleus.  A similar mechanism may underlie temperature-dependent decay in surface tension coefficients in simple liquids, and may also play a role in viscosity and surface tension variations in, e.g., polar and ionic liquids. 

\section*{Phonons and self-diffusion}
In pursuing our objective of developing a picture of single molecule dynamics, within the framework of Langevin's equation (\ref{langevineqnsimple}), we confront two additional, connected questions: a) What is the physical origin of the random force. $ \mathbf{F_R} \left( t \right)$? Typically, $ \mathbf{F_R} \left( t \right)$ is treated as a statistical entity, endowed with physically reasonable statistical properties \cite{zwanzig, kubo}. In liquid-state dynamics problems, this mathematical approach reflects our poor physical understanding of $ \mathbf{F_R} \left( t \right).$ b) What is the origin of self-diffusion,  i.e., the random, thermally-driven motion of individual molecules though a liquid? Since $ \mathbf{F_R} \left( t \right)$ drives self-diffusion, answering either question offers insight into both. 

There are two significant experimental clues: a) A series of experiments, carried out in the 1970's \cite{kinetictheoryX}, measured the self-diffusion coefficient, $ D_s = D_s \left( T, P \right) ,$ in liquid Ar, Kr, and Xe, over a range of temperatures and a series of fixed pressures, and lead to a (dimensionless) correlation of the following form:  
\begin{equation}\label{dsexp}
D_s^* \left( T^* , P^*  \right) =  1.1 \exp{ \left( 0.16 P^* \right)} \exp{ \left[ - \frac{2.39 + 0.23 P^*}{T^*} \right]}
\end{equation}
where
$ D_s^* = D_s / \sqrt{ \epsilon \sigma^2 / M} , $ $ T^* =  T / \left( \epsilon/k_B \right), $ and $ P^*= P / \left( \epsilon / \sigma^ 3 \right) .$ 
b) More recently, Bolmatov, Brazhkin and Trachenko \cite{bolmatov, phononliqreview} presented strong evidence that temperature-dependent specific heats, in a large family of liquids, reflect existence of dominant, solid-state-like, equilibrium phonon modes. 

Consider the solid-like dynamics of N-molecule liquid-state systems, over time scales ranging from the Frenkel to the dispersion scales,  $ \tau_{F} = 2 \pi/ \omega_F = O \left( 10^{-14} \ \mathrm{s} \right) $ to $ \tau_d = 2 \pi/ \omega_d = O \left( 10^{-16} \ \mathrm{s} \right) .$  Using a normal mode analysis - as in solid-state systems - under the assumption that individual molecular oscillations remain small enough to approximate intermolecular potentials as quadratic in the oscillation amplitude, 3N independent, vibrational, i.e., phonon modes are determined. The principal limitation of this model centers on ignoring the random hops of individual molecules. However, based on the arguments and results below, it appears that molecular hopping, over the spectral range  $ \omega_F  < \omega < \omega_{d} , $ is limited to a narrow, low-frequency band of frequencies near the solid state limit: $ \omega \approx \omega_F .$

On time scales on the order of, and shorter than $ \tau_F , $ and in the absence of single-molecule-scale external perturbations - like those produced by short wavelength neutron scattering beams - individual nuclei undergo small displacements, on the order of a small fraction of $ \sigma . $  Under these conditions, and in terms of the Langevin model - see Table 2 - the friction force can be neglected and the dynamics of individual molecules modeled using:
\begin{equation}\label{dseqnstart}
M \frac{d \mathbf{v}}{d t} = -M \sum_{i=1} \omega_i^2 \int_{0}^t \tilde{\mathbf{v}}_i \left( t , \tau \right) d \tau
\end{equation}

This equation states that on these time scales, individual molecules are subject to the summed effect of all phonon modes extant over the spectrum, $ \omega_F \leq \omega \leq \omega_d .$ Specifically, the phonon mode having frequency $ \omega_i , $ induces an instantaneous nuclear velocity $ \tilde{\mathbf{v}}_i \left( \tau, \omega_i \right) ,$ which, in turn, produces a nuclear displacement - over the small time interval $ \left[0,t \right] $ -  of $ \Delta \mathbf{x}_i \left( \tau, \omega_i \right) = \int_0^t \tilde{\mathbf{v}}_i \left( \tau , \omega_i \right) d \tau . $ Since phonon modes are independent, the small displacements, $ \Delta \mathbf{x}_i \left( \tau, \omega_i \right) ,$ are likewise. Thus, since $ M \omega_i^2 \Delta \mathbf{x}_i \left( t \right) $ corresponds to the $ i^{th} $ instantaneous spring force on the molecule, the sum of the random phonon forces corresponds to the right side of (\ref{dsexp}). 

\subsection*{Derivation of the self-diffusion coefficient, $ D_s $}
From Note f), section 3, on the solid-state-like time scale, $ \tau_d \leq t \leq \tau_F ,$ the equation describing nuclear motion, driven by the $ j^{th} $ phonon mode, is given by:
\begin{equation}\label{phonondyneqn}
\frac{ d \mathbf{\tilde{v}_j} \left( t; \omega_j \right)}{dt} = - \omega_j^2 \int_0^t \mathbf{\tilde{v}_j} \left( t' ; \omega_j \right) dt' \approx - \omega_j^2 \mathbf{\tilde{v}_j} \left( t; \omega_j \ \right) t
\end{equation}
Solving this leads to
\begin{equation}\label{vtildesoln}
\mathbf{\tilde{v}_j} \left( t ; \omega_j \right)  = \mathbf{\tilde{v}_j} \left( 0; \omega_j \right) \exp \left[ \frac{ - \omega_j^2 t^2}{2} \right] 
\end{equation}
Thus, the instantaneous velocity of the nucleus corresponds to the superposition of phonon-induced velocity contributions:
\begin{equation}\label{velocityduetophonons}
\mathbf{v} \left( t \right) = \sum_{j}  \mathbf{\tilde{v}_j} \left( t, \omega_j \right)
\end{equation}
so that the dot product, $ \mathbf{v} \left( t \right) \cdot \mathbf{v} \left( 0 \right),$ is given by:
\begin{equation}\label{velocitydotproduct}
\mathbf{v} \left( t \right) \cdot \mathbf{v} \left( 0 \right) = \sum_{j}  \mathbf{\tilde{v}_j} \left( t, \omega_j \right) \cdot \sum_{i}  \mathbf{\tilde{v}_i} \left( 0, \omega_i \right)
\end{equation}
Due to the independence of phonon modes,
\begin{equation}\label{velocitydotproductfinal1}
\langle \mathbf{v} \left( t \right) \cdot \mathbf{v} \left( 0 \right) \rangle = \langle \sum_{j}  {\tilde{v}^2_j} \left( 0; \omega_j \right) \exp \left[ \frac{ - \omega_j^2 t^2}{2} \right] \rangle
\end{equation}
where $ {\tilde{v}^2_j} \left( 0; \omega_j \right)  =   \mathbf{\tilde{v}_j} \left( 0, \omega_j \right) \cdot \mathbf{\tilde{v}_j} \left( 0, \omega_j \right) .$

The self-diffusion coefficient, $ D_s ,$ is given by the integrated velocity autocorrelation function:
\begin{equation}\label{dseqnapp1}
D_s = \int_0^{\infty} \langle \mathbf{v} \left( t \right) \cdot \mathbf{v} \left( 0 \right) \rangle \ dt 
\end{equation}
or
\begin{equation}\label{dseqnapp2}
D_s = \int_0^{\infty}  \Big \langle \sum_j {\tilde{v}^2_j} \left( 0; \omega_j \right) \exp \left[ \frac{ - \omega_j^2 t^2}{2} \right] \Big  \rangle  \ dt
\end{equation}
Integrating then gives:
\begin{equation}\label{dseqnapp3}
D_s = \sqrt{\frac{\pi}{2}} \Big \langle  \sum_j \frac{1}{\omega_j} {\tilde{v}^2_j} \left( 0; \omega_j \right) \Big \rangle
\end{equation}

In order to evaluate the equilibrium average: i) recall that within a given volume, $ V, $ the average number of phonons having frequency $ \omega $ is given by \cite{pathria, reif}
\begin{equation}\label{nomegaequil}
\langle n_{\omega} \rangle = \frac{1}{\exp {\beta \tilde{h} \omega} -1}
\end{equation}
ii) at any location in $ V , $ assume that the wave vector associated with each mode, over the ensemble, is isotropically oriented, and
iii) due to the nominally continuous distribution of modes, move to a continuum representation of the average in (\ref{dseqnapp3}):
\begin{equation}\label{dsint1}
D_s = \sqrt{\frac{\pi}{2}} \int_{\omega_F}^{\omega_d} \frac{ g \left( \omega \right)}{\exp {\beta \tilde{h} \omega} -1} \frac{\tilde{v}^2 \left( 0 ; \omega \right)}{ \omega} \  dw
\end{equation}
where $ g \left( \omega \right) $ is the density of modes \textit{driving self-diffusion}.
Finally, in order to arrive at a theoretical $ D_s $ having the 
same generic structure as the empirical $ D_s $ in (\ref{dsexp}),
we assume that the density of modes driving self-diffusion is clustered around a critical frequency, $ \omega_c : $
\begin{equation}\label{gomega}
g \left( \omega \right) = \delta \left( \omega - \omega_c \right)
\end{equation}
As described below, this assumption leads to a detailed, physically consistent explanation of phonon-driven self-diffusion in simple, nonpolar, nonmetallic liquids. 

Using (\ref{gomega}) in (\ref{dsint1}), approximating $ \exp {\beta \tilde{h} \omega} -1 $ as $ \exp {\beta \tilde{h} \omega} ,$ and nondimensionalizing using $ D_s^* = D_s / \sqrt{ \epsilon \sigma^2 / M} , $ $ T^* =  T / \left( \epsilon/k_B \right), $ and $ P^*= P / \left( \epsilon / \sigma^ 3 \right) , $ finally leads to: 
\begin{equation}\label{dsfinal}
D_{s, \alpha}^* \left( T^* , P^* \right) = \frac{ \langle \tilde{v}_{c,\alpha}^2  \rangle 
\sqrt{ \pi /2} }{ \omega_{c, \alpha} \left( \frac{ \epsilon_{\alpha} \sigma_{\alpha}^2}{M_{\alpha}} \right)^{1/2} } \exp{ \left[ -\frac{\tilde{h} \omega_{c, \alpha} / \epsilon_{\alpha}}{T_{\alpha}^*} \right]} 
\end{equation}
where $ \alpha $ denotes either Ar, Kr, or Xe, and where two
undetermined, pressure-dependent parameters, $ \langle \tilde{v}^2 \left( 0, \omega_c \right) \rangle $ and $ \omega_{c, \alpha} ,$ appear.  The first, 
\begin{equation}\label{velphononc}
\langle \tilde{v}_{c,\alpha}^2 \rangle = \langle \tilde{v}^2 \left( 0, \omega_c ; P^* \right) \rangle
\end{equation}
 is the phonon-induced, ensemble averaged, pressure-dependent, squared velocity of the molecule, evaluated at the critical phonon frequency, 
\begin{equation}\label{omegac}
\omega_{c,\alpha} = \omega_{c,\alpha} \left( P^* \right) 
\end{equation}
where $ \omega_c $ is the frequency that induces significant, single-atom-scale, random jumps, i.e., self-diffusion.  The physical meaning of these parameters is explored in the next section.
Note, approximating $ \exp {\beta \tilde{h} \omega} -1 $ as $ \exp {\beta \tilde{h} \omega} ,$ - again, introduced in order to arrive at a theoretical $ D_s $ having the same form as (\ref{dsexp}) - is based on the fact that, in liquid Ar, Kr, and Xe, $ \exp {\beta \tilde{h} \omega} = O \left( 10 \right) .$

\subsection*{Phonon-induced hopping speeds and critical frequencies; comparisons with experimental self-diffusion coefficients}
In order to determine $ \langle \tilde{v}_{c,\alpha}^2 \rangle $ and $ \omega_{c,\alpha} , $ we use the experimental correlation \cite{kinetictheoryX} (\ref{dsexp}), 
leading to 
\begin{equation}\label{omegaeqn}
\omega_{c, \alpha } \left( P^* \right) = \left( 2.39 + 0.23 P^* \right) \epsilon_{\alpha} / \tilde{h}    
\end{equation}
 and 
\begin{equation}\label{vsquaredeqn}
\langle \tilde{v}_{c,\alpha}^2 \left( P^* \right) \rangle = 1.1 \cdot \omega_{c, \alpha} \left( \frac{ \epsilon_{\alpha} \sigma_{\alpha}^2}{M_{\alpha}} \right)^{1/2} \sqrt{2/ \pi} \exp{ \left( 0.16 \cdot P^* \right) }
\end{equation}

Comparisons of temperature- and pressure-dependent self-diffusion coefficients, $ D_s^* \left( T^*, P^* \right) , $ predicted by the phonon-based model, (\ref{dsfinal}), with experimental measurements \cite{kinetictheoryX} in liquid Ar, Kr, and Xe, are shown in Figs. \ref{argonselfdiffnfig}, \ref{kryptonselfdiffnfig}, and \ref{xenonselfdiffnfig}.  Pressure-dependent magnitudes of the critical phonon frequency, $ \omega_{c, \alpha },$ driving self-diffusion, and the root mean square atomic speed, $ \sqrt{ \langle \tilde{v}_{c,\alpha}^2 \left( P^* \right) \rangle } , $ induced by these critical phonons, are listed in Table \ref{tab:table2}. 

As a preliminary consistency check on this picture of phonon-driven self-diffusion, leading to the  semi-empirical expressions for the characteristic hopping frequency, $\omega_{c,\alpha} ,$ and speed of hopping molecules in $\langle \tilde{v}_{c,\alpha}^2 \rangle , $ (\ref{omegaeqn}) and (\ref{vsquaredeqn}), respectively, we note the following:


\begin{figure}[!tbhp]
  \centering
    \includegraphics[width=1.0\linewidth, trim = {0 2.5cm 0 1cm}]{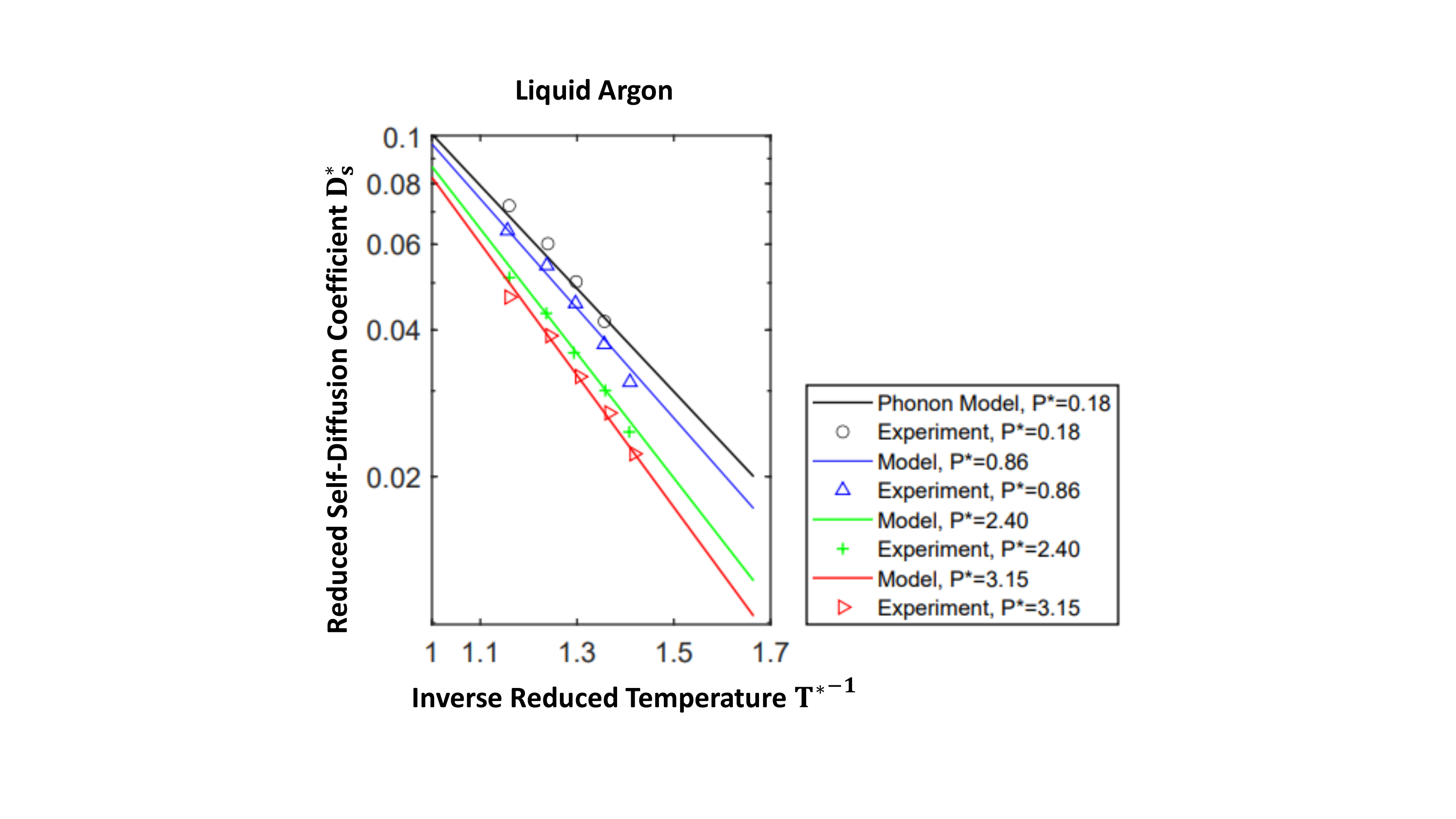}
    \caption{Temperature-dependent self-diffusion coefficient of liquid argon. Over the spectrum of frequencies available to a liquid state system, $ 0 \leq \omega \leq \omega_d , $ the phonon self-diffusion model: i) idealizes the band from the Frenkel frequency, $ \omega_F = \mathrm{O} \left( 10^{14} \ \mathrm{s^{-1}} \right), $ to the dispersion frequency, $ \omega_d = \mathrm{O} \left( 10^{16} \ \mathrm{s^{-1}} \right), $ as corresponding to solid-state-like dynamics, ii) assumes that on $ \omega_F \leq \omega \leq \omega_d , $ individual molecules undergo small amplitude, harmonic vibrations about fixed positions, and iii) thus allows a normal mode analysis of the solid-like dynamics. In order to capture the observed temperature dependence of $ D_s^* $ \cite{kinetictheoryX}, it is necessary to assume that the band of phonon frequencies driving self-diffusive, single molecule random hops is concentrated near the low end of the solid state spectrum, $ \omega \approx \omega_F,$ idealized as a delta function in (\ref{gomega}). The nondimensional definition of $ D_s^* $ is given following (\ref{dsexp}).} 
    \label{argonselfdiffnfig}
\end{figure}

\begin{figure}[!tbhp]
  \centering
     \includegraphics[width=1.0\linewidth]{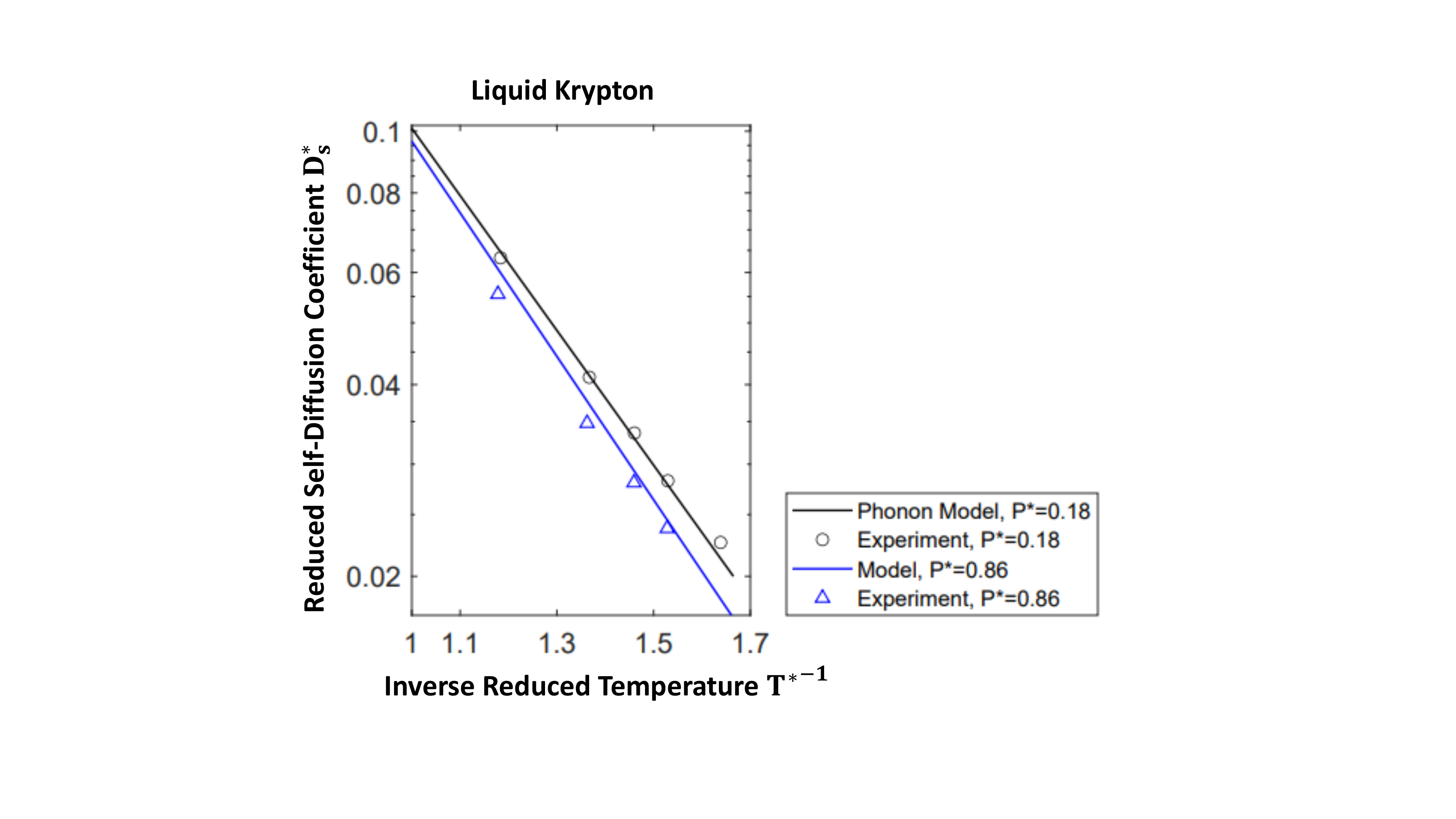}
    \caption{Temperature-dependent self-diffusion coefficient of liquid krypton. See the caption to Fig. \ref{argonselfdiffnfig} for a description of the phonon model of self-diffusion.}
    \label{kryptonselfdiffnfig}
\end{figure}

\begin{figure}

\includegraphics[width=1.0\linewidth, trim = {0 2cm 0 1.5cm}]{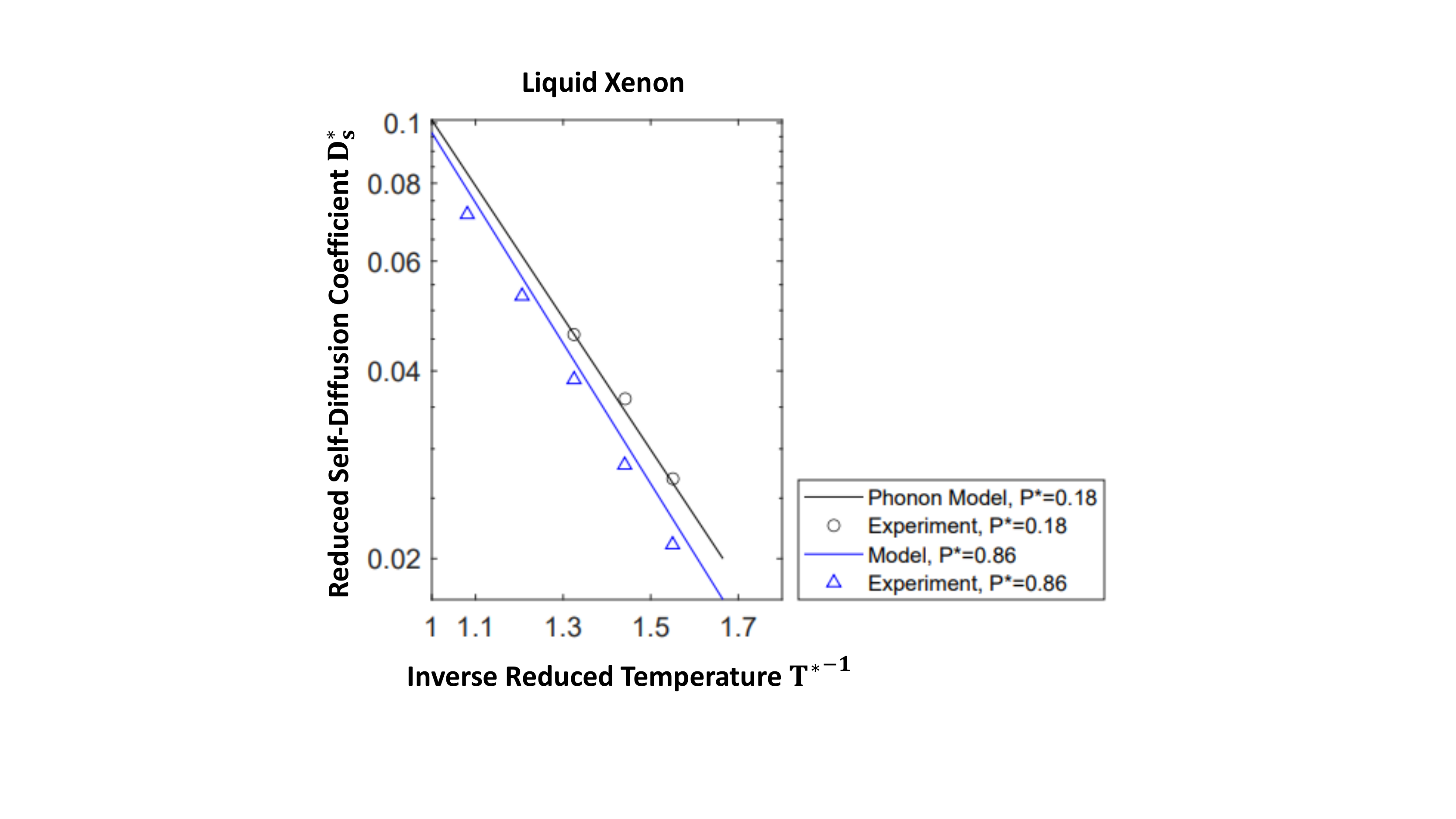}
\caption{Self-diffusion coefficient for liquid Xenon. See the caption to Fig. \ref{argonselfdiffnfig} for a description of the phonon model of self-diffusion.}
\label{xenonselfdiffnfig}
\label{fig:wrapfig}
\end{figure}

\begin{center}
\begin{table*}[!bhp]
\begin{adjustwidth}{-1.5in}{0in}
\begin{tabular}{|c|c|c|c|c|}
\hline
\hline
\multicolumn{5} {|c|} {}  \\
\multicolumn{5} {|c|} {\Large{Phonon Induced Hopping Speed and Critical Frequency}} \\
\multicolumn{5} {|c|} {} \\
\hline
\hline
 Reduced Pressure & P*=0.18 & P*=0.86 & P*=2.4 & P*=3.15 \\
\hline
Specie & $ \sqrt {\langle{v}_{c}^2 \rangle} $ \hspace{.5 cm}    $ \omega_c \times 10^{13} $ & $ \sqrt {\langle{v}_{c}^2 \rangle} $ \hspace{.5 cm}    $ \omega_c \times 10^{13} $  & $ \sqrt {{v}_{c}^2 } $ \hspace{.5 cm}    $ \omega_c \times 10^{13} $ & $ \sqrt {{v}_{c}^2 } $ \hspace{.5 cm}   $ \omega_c \times 10^{13} $ \\
 & (m/s) \hspace{ .5cm}   $ \mathrm{s^{-1}} $  & (m/s) \hspace{ .5cm}   $ \mathrm{ s^{-1}} $ &
(m/s) \hspace{ .5cm}   $ \mathrm{s^{-1}} $ &  (m/s) \hspace{ .5cm}   $ \mathrm{s^{-1}} $  \\
\hline
Ar & 1340 \hspace{ .5cm} 3.70 & 1470 \hspace{ .5cm} 3.94 & 1780 \hspace{ .5cm} 4.48 & 1950 \hspace{ .5cm} 4.74 \\
\hline
 Kr & 1660 \hspace{ .5cm} 6.06 & 1813 \hspace{ .5cm} 6.45 & NM & NM
 \\
\hline
Xe & 1814 \hspace{ .5cm} 7.30 & 1980 \hspace{ .5cm} 7.77 & NM & NM  \\
\hline    
\end{tabular}
\caption{\label{tab:table2}According to the proposed model of phonon-driven self-diffusion, over the portion of the frequency spectrum where liquid dynamics are solid-like, $ \omega_F \leq \omega \leq \omega_d , $ the instantaneous velocity of any given nucleus is determined by superposition of $ 3N $ independent, phonon-induced velocity contributions: $ \mathbf{v} \left( t \right) = \sum_{j}  \mathbf{\tilde{v}_j} \left( t, \omega_j \right) . $ Based on this correspondence and the assumption, (\ref{gomega}), that self-diffusive hops are produced by a narrow band of phonon modes centered near a critical frequency, $ \omega_c ,$ we identify $ \sqrt {\langle{v}_{c}^2 \rangle}  $ as the hopping speed. NM = not measured.}
\end{adjustwidth}
\end{table*}
\end{center}

\vspace{0.05cm} 

\noindent a) For small to moderate reduced pressures, $ P^* = O \left( 1 \right) , $ (\ref{vsquaredeqn}) leads to the following approximate equality:
\begin{equation}\label{dsapprox}
\frac{\langle \tilde{v}_{c,\alpha}^2 \rangle }{ \omega_{c, \alpha}} \approx  \left( \frac{ \epsilon_{\alpha} \sigma_{\alpha}^2}{M_{\alpha}} \right)^{1/2} \left( = D_{so, \alpha} \right)
\end{equation}
where $ D_{so, \alpha} $ is the scale of the self-diffusion coefficient.  By contrast, as a check on the steps leading from insertion of the integrated single molecule dynamics equation, (\ref{dseqnstart}), to the expression for $ D_s ,  $ 
written in the form:
\begin{equation}\label{finaldseqn}
D_s = \int_0^{\infty} \left[ \int_{\omega_c}^{\omega_D} \frac{g \left( \omega \right) }{ \left( \exp {\beta \tilde{h} \omega}-1 \right)} f \left( \omega \right) d \omega \right] dt
\end{equation}
the left side of (\ref{dsapprox}) can be obtained by starting with the definition, $ D_{s, \alpha} = \int_0^{\infty} \langle \mathbf{v} \left( t' \right) \cdot \mathbf{v} \left( 0 \right) \rangle dt' , $ and replacing the upper limit with the characteristic time scale for single-molecule hops, $ \tau_{hop, \alpha} = \omega_{c, \alpha}^{-1} ,$ where the latter captures the assumed delta-function density of hop-inducing phonons near $ \omega_{c, \alpha} ,$ (\ref{gomega}). This leads to $ D_{s, \alpha} = \langle \mathbf{v} \left( 0 \right) \cdot \mathbf{v} \left( 0 \right) \rangle \cdot \omega_{c , \alpha }^{-1} = \langle \tilde{v}_{c, \alpha}^2 \rangle \cdot \omega_{c , \alpha }^{-1} . $ 

\vspace{0.3cm}

\noindent b) For all three liquids, Ar, Kr and Xe, estimated critical phonon frequencies, $ \omega_{c, \alpha } $ - which we interpret as the characteristic hopping frequency - lie well within the range of frequencies, $ \omega_{d, \alpha} >  \omega_{c, \alpha} > \omega_{F, \alpha} , $
where these liquids maintain solid-like properties \cite{phononliqreview}.  Estimated $ {\omega_{c, \alpha}}'s $ are approximately six times higher than estimated Frenkel frequencies \cite{bolmatov, phononliqreview}, $ \omega_{c, \alpha } \approx 6 \omega_{F, \alpha} = 12 \pi G_{\infty, \alpha} / \mu_{\alpha} ,$ and approximately two orders of magnitude smaller than characteristic dispersion frequencies, $ \omega_{d, \alpha} ,$
where $ G_{\infty , \alpha} $ and $ \nu_{\alpha} $ are, respectively, the high-frequency shear modulus \cite{boonyip, phononliqreview, bolmatov} and dynamic viscosity of specie $ \alpha ,$ and where magnitudes of $ G_{\infty, \alpha} $ are obtained from \cite{bolmatov}, and magnitudes of $ \nu_{\alpha} $ are given in Table S.1 in the Appendix.  

\vspace{0.3cm}

\noindent c) Magnitudes of molecular hopping speeds, $ \sqrt{ \langle \tilde{v}_{c,\alpha}^2 \rangle }, $ 
exceed, by roughly a factor of two, both the longitudinal liquid-state sound speed \cite{dyre2006}, $ a_{liq} = \sqrt{ K / \rho } ,$ and the slightly faster longitudinal solid-state sound speed, $ a_{solid} = \sqrt{ a_L^2 + 4/3 a_S } ,$ where $ K $
is the bulk modulus and $ a_S = \sqrt{ G_{\infty} / \rho }  $ is the shear (transverse) wave speed. Thus, average atomic hopping speeds are well in excess of characteristic liquid- and solid-state sound speeds. Equivalently, from (\ref{omegaeqn}), the hop-inducing phonon energy, $ \tilde{h} \omega_{c, \alpha} \left( P^* \right) ,$ is approximately twice the intermolecular energy scale, $ \epsilon_{\alpha} ,$ and increases (linearly) with pressure.


\section*{Electron cloud compression, single molecule hopping, and shear-induced viscosity generation}
We highlight two observations.  First, configuration-averaged instantaneous normal mode (INM) spectra in solid- and liquid-state systems show that, at any instant, a significant fraction of interacting molecular pairs in liquids are in a state of mutual repulsion; in corresponding solids, only a small fraction of pairs are in such states \cite{stratt}. Second, the velocity autocorrelation function (VACF) in Lennard-Jones liquids, e.g., noble gas liquids, is largely determined by the repulsive part of the LJ potential \cite{kushickberne}. 

Combined with the results above, these observations lead to a fairly detailed picture of the solid-state-like phonon modes extant over the frequency band, $ \omega_d \leq \omega \leq \omega_F , $ as well as the mechanism that appears to drive molecular hopping at the lower end of this band. Writing Shrodinger's equation for an individual molecule, $ i \tilde{h} \psi_{, t} +  \tilde{h}^2 \nabla^2 / \left( 2 M \right) = V \psi , $ estimating the scales of the two terms on the left side, where the time scale is taken as the critical hopping frequency, $ t \sim \omega_{c}^{-1} , $ we find that the length-scale on which quantum uncertainty effects are important, $ x_Q = x_{DeBroglie} \sim \sqrt{ \tilde{h} / 2 M \omega } , $ is small relative to the molecular diameter: $ \sigma : $ $ x_Q / \sigma \sim 10^{-2} . $ Thus, at least in the vicinity of $ \omega \sim \omega_c , $ nuclear motion is classical.
%

Next, express the critical phonon-induced hopping speed, $ \sqrt{ \langle  \tilde{v}_{c, \alpha}^2 \rangle } , $ as $ \sqrt{ \langle  \tilde{v}_{c, \alpha}^2 \rangle }  = \omega_c / k_c , $ where $ k_c = 2 \pi / \lambda_c , $ and where $ \lambda_c ,$ the critical phonon wavelength, is on the order of $ 2/3 \sigma .$ Since $ \lambda_c $ corresponds to the largest phonon wavelength, we find that the spectrum of phonon modes, $ \omega_F \leq \omega_i \leq \omega_d , $ corresponds to $ 3N $ independent, small-amplitude oscillations, $ \lambda_i \leq 2 \sigma /3, $ where $ i$ runs from $ 1$ to $ 3N.$ Thus, in contrast to, e.g., crystalline solids, collective, multi-molecule oscillations are nonexistent.

\begin{figure}[!tbhp]
  \centering
    \includegraphics[width=\textwidth, trim = {0cm 1cm 0cm 4cm}]{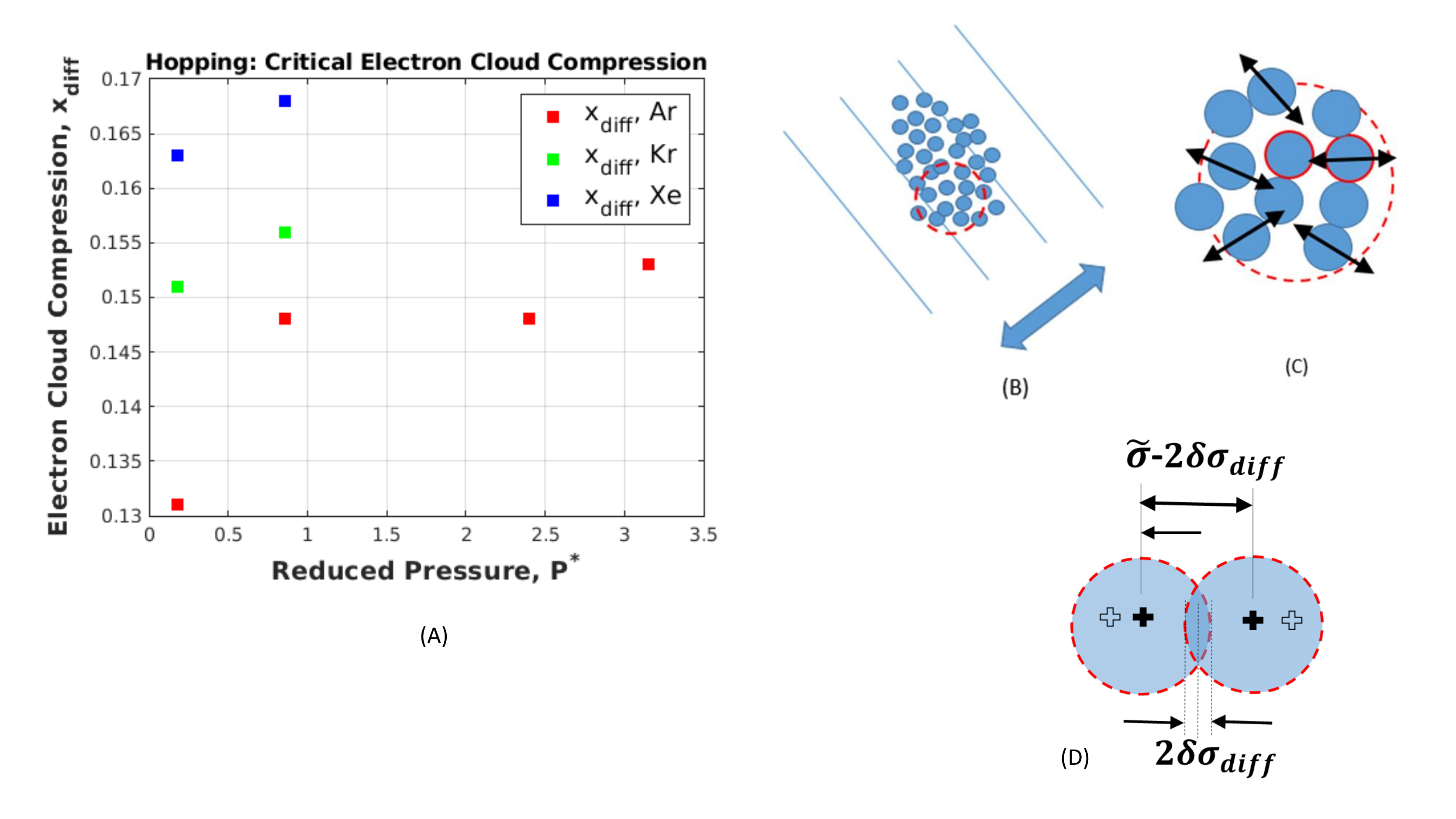}
        \caption{Two observations \cite{stratt, kushickberne} provide important clues concerning the mechanism driving self-diffusion in liquids: i) \cite{stratt} shows that at any instant, a significant fraction of interacting molecular pairs in liquid Ar exist in a state of mutual repulsion, while in the corresponding solid, only a small fraction of pairs are in such states. ii) The velocity autocorrelation function (VACF) in Lennard-Jones liquids, e.g., noble liquids, is largely determined by the repulsive part of the potential \cite{kushickberne}. Viewed in terms of the proposed phonon model of self-diffusion, and given the dominance of pair-wise collisions - see point i) in the final section - these observations suggest that the relatively large single-molecule kinetic energies required for hopping are supplied by collisional compression of adjacent electron clouds. Here, $ x_{diff} = \delta \sigma_{diff} / \tilde{\sigma} ,$ is the estimated relative compression of individual clouds, and $ \tilde{\sigma} = 2^{1/6} \sigma $ is the intermolecular separation minimizing the LJ potential.  An energy conservation argument, leading to (\ref{consenergyhops}), can be used to connect $ \delta \sigma_{diff} , $ to the critical, phonon-induced nuclear velocity, $ \sqrt{\langle v_c^2 \rangle }, $ and the
        critical phonon frequency, $ \omega_c . $ In order to contrast phonon modes, whose wavelengths are all smaller than or approximately equal to $ 2 \sigma / 3 , $ with collective, hydrodynamic modes that emerge on time scales exceeding $ \tau_c = \mathrm{O}\left( 10^{-13} \ \mathrm{s} \right) $ $\Large[$ where solid-like dynamics take place on $ \tau_d = \mathrm{O} \left( 10^{-16} \ \mathrm{s} \right) \leq t \leq \tau_F = \mathrm{O} \left( 10^{-14} \ \mathrm{s} \right) $  $ \Large] ,$ (B) depicts a (short-wavelength, standing) sound wave (blue double arrow). Panel (C) depicts the jitter-like phonon oscillations that are superposed on slower hydrodynamic modes. Panel (D) depicts the electron cloud compression driving self-diffusion.} 
    \label{xdiffxnucfig}
    \end{figure}

The apparent mechanism driving molecular hops is sketched in Fig. \ref{xdiffxnucfig}.  Again, as argued in the final section, point i), on all time scales exceeding $ O \left( \tau_d \right), $ pair-wise intermolecular collisions dominate 3-body and higher-order collisions. Since nuclear motion on the $ \tau_F $ time scale is classical, we can apply the classical version of conservation of energy to the interaction between a fixed target molecule, $ \mathcal{A}, $ and a colliding molecule, $ \mathcal{B} .$ On approach toward $ \mathcal{A}, $ $ \mathcal{B} $ is assumed to have sufficient (relative) kinetic energy and (relative) momentum to allow $ \mathcal{A} $ and $ \mathcal{B} $ to enter a repulsive state. Applying conservation of energy to $ \mathcal{B} ,$ from the instant when maximum electron cloud compression occurs - and the relative velocity of $ \mathcal{B} $ is $ \mathbf{0} $ - to the instant when the intermolecular separation, $ r_{\mathcal{AB}} , $ equals the LJ potential minimizing separation, $ \tilde{\sigma} = 2^{1/6} \sigma ,$ we obtain: 
\begin{equation}\label{consenergyhops}
\frac{M}{2} \left[ \langle v_c^2 \rangle - \langle v_i^2 \rangle \right] \approx - \epsilon \int_{r_c}^{\tilde{\sigma}} \frac{\partial}{\partial r} \left( \frac{\tilde{\sigma}}{r} \right)^{1/12} dr
\end{equation}
where the intermolecular potential is dominated by repulsion, and where
the equation represents the ensemble average dynamics of $ \mathcal{A} $ and 
$ \mathcal{B} $ for a single collision.  Using $ \langle v_i^2 \rangle = 0 ,$ as well as the relationship $ \delta \sigma_{diff} = \left( \tilde{\sigma} - r_c \right) /2 ,$ then leads to an estimate for the fractional electron cloud compression, 
$ x_{diff} ,$ that produces single molecule hops:
\begin{equation}\label{fracecompn}
x_{diff} = \frac{\delta \sigma_{diff}}{\tilde{\sigma}} \approx \frac{1}{2} \left[ 1 - \left( 1 + \frac{M \langle v_c^2 \rangle }{2 \epsilon} \right)^{-1/12} \right]
\end{equation}
Estimated, pressure-dependent magnitudes of $ x_{diff} $ for Liquid Ar, Kr and Xe, are plotted in Fig. \ref{xdiffxnucfig}.

\begin{figure}[!tbhp]
  \centering
     \includegraphics[width=1.0\textwidth, trim = {0cm 3.5cm 0cm 1.5cm}]{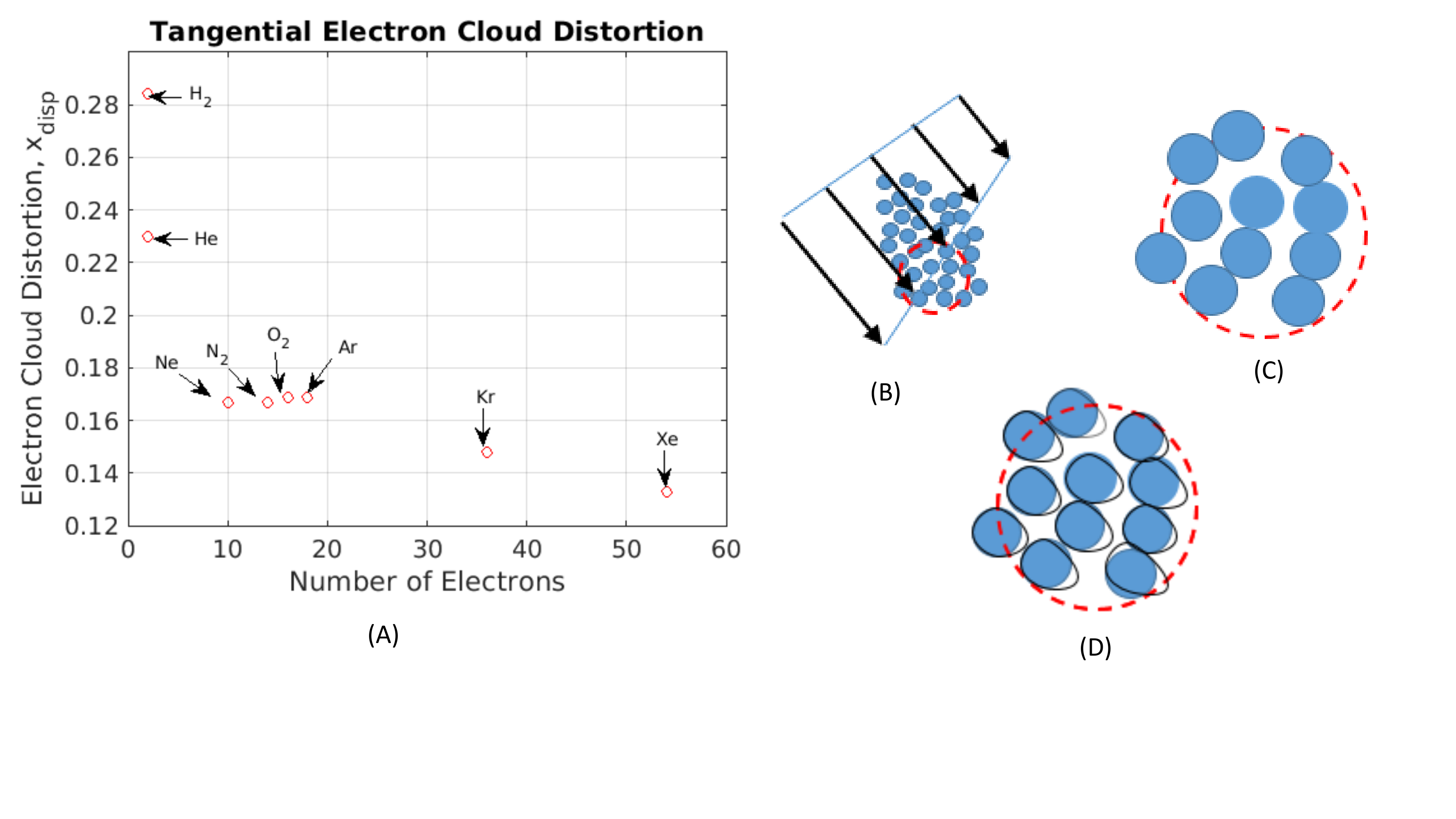}
    \caption{Distortion of electron clouds appears to play a dominant role in both emergence of single-molecule-scale viscosity and resistive viscous forces, as well as in repulsion-driven single molecule hopping.  Intuitively, we anticipate that the magnitudes of cloud distortion associated with each process, $ \delta \sigma_{disp} $ and $ \delta \sigma_{diff} ,$ respectively, should be roughly of the same magnitude.  In order to test this idea, and as a consistency check on the proposed models of dispersion-induced viscosity generation and phonon-induced self-diffusion, we estimate the viscosity-generating cloud distortion, $ \delta \sigma_{disp} , $ using an expression equivalent to (\ref{polareqn}) for the polarization \cite{hirschfelder}: $ \alpha = 4 n \langle r_1^2 \rangle^2 / \left(9 a_o \right) ,$ where $ n $ is the number of electrons in the molecule, and $ \langle r_1^2 \rangle / $ is the mean squared collision-induced displacement of any of the (indistinguishable)  electrons
    occupying the first shell of the molecule. Labeling the quartic root of the latter as $ \delta \sigma_{disp} ,$ and identifying this as the characteristic \textit{tangential} cloud distortion, we calculate the relative tangential  distortion as: $ x_{disp} = \delta \sigma_{disp} / \sigma .$ As shown in (A), outside of He and $ \mathrm{H_2} $ - which, due to small masses, are apparently dominated by quantum collision dynamics -  estimated tangential cloud distortions are approximately of the same magnitiude as those producing self-diffusion, Fig. \ref{xdiffxnucfig}. Thus, while cloud distortions are of comparable magnitude, the \textit{type} of distortion, compressive versus shearing, engages repulsive versus attractive intermolecular forces.  Plate (B) highlights the essential role of microscale (long-time-averaged) shear stresses in driving tangential cloud distortion and resulting viscosity generation. Plate (C) depicts liquid state molecules under local equilibrium conditions (sans phonon jitter).  Nonequilibrium, shear-driven, tangential cloud distortion appears as small white areas at the edge of each molecule in Plate (D).}
    \label{ecloudfig}
\end{figure}

An important consistency check on the proposed pictures of
phonon-driven self-diffusion, embodied by $  x_{diff} $  in (\ref{fracecompn}), as well as dispersion-induced generation of viscosity, can be carried out by combining (\ref{polareqn}) and (\ref{meanelectrondisplacement}) to estimate relative distortions, $ x_{disp} , $ of  electron clouds  accompanying generic (i.e., mostly attractive) intermolecular interactions:
\begin{equation}\label{eclouddistortion2}
x_{disp} = \frac{\delta \sigma }{\sigma} \approx \left[ \frac{9}{4} \frac{ \alpha a_o}{n} \right]^{1/4} \sigma^{-1}
\end{equation}
Note that use of the definition of $ \delta \sigma ,$ given by (\ref{meanelectrondisplacement}), means that we are approximating
the set of level-dependent mean squared electron displacements, $ \overline{r_i^2} , $ as $ \overline{r_1^2} $ \cite{hirschfelder}; thus, plotted magnitudes of $ x_{disp} $ represent slight
underestimates. As shown in Fig. \ref{ecloudfig}, characteristic collision-induced electron cloud distortions for a number of noble and diatomic liquids are approximately of the same magnitude as estimated relative cloud compressions, $ x_{diff} , $ driving self-diffusion.  

Assuming that the former accompanies tangential/glancing interactions - giving rise to single-molecule-scale viscous drag forces, and that the latter characterizes normal/head-on interactions - giving rise to repulsive states and single molecule hops - we arrive at an important, though preliminary picture of the apparent connections between collision-induced electron cloud distortions and single molecule dynamics in nonpolar liquids. See the captions to Figs. \ref{xdiffxnucfig} and \ref{ecloudfig}. More work is required, of course, but these conceptual insights may prove useful.

\section*{Time scale-dependent models of single molecule dynamics}
For nonpolar, spherical atomic liquids like Ar, Kr, and Xe, as well as nonspherical, nonpolar liquids like those examined in \cite{volumestokeseinstein}, we propose that single molecule dynamics can be modeled on three distinct time scales: a) over $ \tau_d \leq t \leq \tau_F , $ dynamics are solid-state-like and either dissipative or not - see below; b) over $ \tau_F \leq t \leq \tau_c , $ dynamics are a mixture of solid- and fluid-like and again, either dissipative or not; and c) for $ t  > \tau_c , $ dynamics are dissipative and fluid-like.  We denote these time scales, respectively, as the solid-like, transitional, and fluid-like regimes.

For clarity, we tabulate in Table 2 the various forms that the single molecule dynamics equation can take.  
The following general points are highlighted:

\vspace{0.2cm}

\noindent \textbf{a) Equation structure:} The proposed equations are physically explicit versions of the memory-free and generalized Langevin equation \cite{boonyip, forster, kubo}:
\begin{equation}\label{langevineqngeneric}
M \frac{ d \mathbf{v} \left( t \right)}{dt} = \mathbf{F_e} \left( t \right) + \mathbf{F_f} \left( t \right) + \mathbf{F_R} \left( t \right)
\end{equation}
where the instantaneous molecular force is decomposed into a time-dependent external force, $ \mathbf{F_e} \left( t \right) , $ necessary for modeling,
e.g., particle scattering problems, a deterministic friction force, $ \mathbf{F_f} \left( t \right) , $  either $ - \int_0^t \kappa \left( t - t' \right) \mathbf{v} \left( t' \right) dt' $ or $ -  3 \pi \sigma_m \mu \mathbf{v} \left( t \right) , $ and a random force, $ \mathbf{F_R} \left( t \right) , $ either determined by the phonon field, $ M \sum_{i=1} \omega_i^2 \int_0^t \tilde{\mathbf{v}}_i \left( t , \tau \right) d \tau , $ or by the thermal motion of surrounding molecules, $ \dot{\mathbf{\eta}} \left(t \right) .$ 

\vspace{0.2cm}

\noindent \textbf{b) Friction force:} The set of equations contrasts the qualitatively distinct dynamics that exist under quiescent conditions, when the continuum-scale liquid flows or is stationary, versus the highly dynamic state extant when the target molecule interacts directly with, or lies near an externally introduced particle.  In the first case, based on the observation that small molecules follow a slightly modified version of the Stokes-Einstein drag force law \cite{volumestokeseinstein}, we assume that the friction force can be expressed as $ -  3 \pi \sigma_m \mu \mathbf{v} \left( t \right) , $ where again, $ \sigma_m $ is an effective, shape-dependent molecular diameter.  This assumption, in turn, assumes that the dynamic viscosity, $ \mu , $ emerges on time scales that are long relative to the fast disperion time scale, $ \tau_d , $ but short relative to the solid-liquid cross-over time scale, $ \tau_F . $ Based on the observation that temperature-dependent viscosities are well-predicted by the above model incorporating fast-acting
dispersion forces, this appears to be a reasonable assumption. Under conditions where, e.g., scattering particles interact with or near the target molecule, numerous experimental observations show that the friction force is history dependent \cite{boonyip, lovesey2, forster, bernepecora}.  

\vspace{0.2cm}

\noindent \textbf{c) Connecting the phonon force to the Brownian force:} Kubo's analysis \cite{kubo} can be adapted to show explicitly how the phonon force on $ \tau_d \leq t \leq \tau_F $ can be represented as a Brownian force on $ t \geq \tau_c : $  i) Express the instantaneous phonon force (in any of three orthogonal directions) as $ F_{phonon} \left(t \right) = M \sum_{i=1}^n \omega_i^2 \Delta x_i , $ where $ \Delta x_i \left( t \right) = \int_0^t \tilde{\mathbf{v}}_i \left( t , \tau \right) d \tau , $ is the nuclear displacement produced by phonon mode $ i, $ and $ \omega_i $ the $ \mathrm{i^{th}} $ normal mode frequency; ii) recognize, by (normal mode) construction, that on $ \tau_d \leq t \leq \tau_F ,$ all $ \Delta x_i ' \mathrm{s} $ are zero-mean, independent random displacements; iii) define the sum of displacement variances as $ s_n^2 = \sum_{i=1}^n \sigma_i^2 ,$ where $ \sigma_i^2 = \langle \Delta x_i^2 \rangle $ is the  $ \mathrm{i^{th}} $ variance; iv) focusing on time scales on the order of $ \tau_c $ and longer, define a random variable $ Y_n \left( t \right) = F_{phonon} \left( t \right)  / s_n ,$ where $ t = \mathrm{O} \left( \tau_c \right) ; $ v)  allow $ n $ to become large, which corresponds to binning all of the random phonon forces acting on $ \tau_d \leq t \leq \tau_F ;$ vi) by the central limit theorem, the probability density for $ Y_n \left( t \right) $ approaches a Gaussian density, $ p \left( Y \left(t \right)  \right) \rightarrow \frac{1}{\sqrt{2 \pi} } \exp { \frac{ -Y^2 }{2}} ;$ and vii) on $ t = \mathrm{O} \left( \tau_c \right) , $ argue that phonon force components (in each of three orthogonal directions) are delta-correlated,
$ \langle F_{phonon} \left( t \right) F_{phonon} \left( t' \right) \rangle = F_o^2 \delta \left( t - t' \right) , $ where $ F_o^2 $ is  the force intensity. In Table 2, this guassian, delta-correlated remnant of the phonon force is labeled as $ \mathbf{\dot{\eta}} \left( t \right) . $

\vspace{0.2cm}

\noindent \textbf{d) The random force, $ \mathbf{F_R} \left( t \right):$} For the solid-like regime, the arguments from the previous section provide, we believe, substantial support for expressing the random force as $ \mathbf{F_R} \left( t \right) = -M \sum_{i=1} \omega_i^2 \int_0^t \tilde{\mathbf{v}}_i \left( t , \tau \right) d \tau .  $ For the fluid-like regime, the fact that the modified Stokes-Einstein relation, (\ref{stokeseinstein}) holds for a large family of molecules \cite{volumestokeseinstein}, where again (\ref{stokeseinstein}) is derivable from the memory-free Langevin equation (\ref{langevineqnsimple}) \cite{chandrasekhar}, suggests that $ \mathbf{F_R} \left( t \right) = \mathbf{\dot{\eta}} \left( t \right) .$ 
Proposing a reasonable form of $ \mathbf{F_R} \left( t \right) $ over the transition regime, $ \tau_F \leq t \leq \tau_c , $ remains problematic at this point, however.  A mathematically simple assumption,
which may not be physically valid, would model $ \mathbf{F_R} \left( t \right) $ as a linear superposition of $ -\sum_{i=1} \omega_i^2 \int_0^t \tilde{\mathbf{v}}_i \left( t , \tau \right) d \tau $ and $ \mathbf{\dot{\eta}} \left( t \right) . $ This is an open question, however. 

\vspace{0.2cm}

\noindent \textbf{e) The external force:} An external force term only appears for problems in which the spatial scale of the external agent, e.g., a scattering particle or a high-energy photon source (having wavelength on the order of $ \sigma $ or smaller), is on the order of the molecular diameter, $ \sigma . $ To account for such forces, a quantum mechanical model of the interaction is typically required; see, e.g.,  \cite{lovesey2, bernepecora}.  

\vspace{0.2cm}

\noindent \textbf{f) Physical meaning of the phonon force:} Over the solid-state-like time scale, $ \tau_d \leq t \leq \tau_F , $ under conditions where molecule-scale external forcing is absent, the phonon field determines: i) each  molecule's instantaneous velocity, $ \mathbf{v} \left( t \right) = \sum_{i}  \mathbf{\tilde{v}_j} \left( t, \omega_j \right) , $ as well as ii) the instantaneous random force, $ \mathbf{F_R} \left( t \right)  = - M \sum_j \omega_j^2  \Delta \mathbf{ \tilde{x}_j} \left( t \right) = - M \sum_j \omega_j^2 \int_0^t \mathbf{\tilde{v}_j} \left( t', \omega_j \right) dt' . $ Thus, the dynamics of individual nuclei: i) can be decomposed into individual contributions produced by each phonon mode: $ M \mathbf{\dot{\tilde{v}}_j} = - M \omega_j^2 \int_0^t \mathbf{\tilde{v}_j} \left( t', \omega_j \right) dt' ,$ or ii) taken as the resultant of these modes: $ M \mathbf{\dot{{v}}} = - M \sum_j \omega_j^2 \int_0^t \mathbf{\tilde{v}_j} \left( t', \omega_j \right) dt' ,$ where time derivatives, denoted by dots, are taken with respect $ t ,$ on the solid state time scale. 

\vspace{0.2cm}

\noindent \textbf{g) On the weak coupling between continuum scale flow and microscale dynamics:} A scaling argument shows that only under extreme circumstances can continuum flow fields produce non-negligible microscale nonequilibrium mass, momentum and energy currents. Consider, for example, turbulent flow over a mirror-smooth surface (having asperities on the order of, say, $ 10^{-9} \ \mathrm{m} ).$ Taking the ratio of the maximum continuum-scale viscous shear stress, evaluated at the surface, $ \tau_{cont} \approx 0.02 \rho U_{\infty}^2 Re_{\delta}^{-1/4} $ \cite{schlichting}, to the characteristic molecular-scale shear stress, $ \tau_{molec} \approx 10 \mu a / \sigma , $ leads to $ \tau_{cont} / \tau_{molec} \sim 0.002 U_{\infty} Ma \sigma ,  $ where $ Re_{\delta} = \rho U_{\infty} \delta / \mu \sim 1 $ is the Reynolds number associated with a turbulent boundary layer of thickness, $ \delta , $ $ U_{\infty} $ is the speed of the flow external to the boundary layer, and $ Ma = U_{\infty} / a ,$ is the associated Mach number. Here, $ \tau_{molec} ,$ which is determined by the transverse momentum current \cite{kubo, boonyip}, is most easily estimated using the Stoke's drag law, $ F_{drag} \approx 3 \pi \sigma \mu a $ \cite{panton},  where the molecular speed is approximated as the sound speed, $ a .$ Using the Mach number magnitude, $ Ma \sim 0.3 , $ separating nominally incompressible and compressible flow, leads to the condition: $ \tau_{cont} \sim \tau_{molec} $ when $ U_{\infty} \sigma / nu \sim 10^3 .$ Due the small magnitude of $ \sigma $ for small molecular species, it is found, for Ar, Kr, and Xe, that $ U_{\infty} $ must be on the order of $ 10^6 \ \mathrm{m/s} ,$ or higher for continuum-scale dynamics to manifest itself in microscale dynamics.

\vspace{0.2cm}

\noindent \textbf{h) Development of short time scale collective dynamics models:} Under the assumption that fast-acting dispersion forces mediate collective dynamics over the elastic, transition, and fluid-like regimes, sum rules \cite{boonyip, forster, kubo, lovesey2} provide a powerful tool for developing hydrodynamic models appropriate to each time scale. Appendix 2 illustrates using a simplified, i.e., non-viscoelastic Navier Stokes model of transition regime collective dynamics. The strategy consists of two steps: 1) propose a model of short-time scale (ensemble average) molecular hydrodynamics, and 2) constrain the model by satisfying sum rules.

\vspace{0.2cm}

\noindent \textbf{i) Dominance of pairwise interactions:} In many single molecule dynamics problems, as well as in derivation of field-based continuum dynamics models \cite{forster, evansmorriss}, it is important to have solid understanding of the relative importance of simultaneous multi-molecule collisions.  At any instant, on any time scale exceeding $ \tau_d , $ consider a target (nonpolar, liquid-state) molecule, $ \mathcal{A,} $ surrounded by a set of neighboring molecules, $ \mathcal{B}_1 , \mathcal{B}_2 , ..., \mathcal{B}_m . $  Since the weak dispersive potential, $ \phi^{(AB_i)} , $ that appears during collision of $ \mathcal{A} $ and $ \mathcal{B}_i , $ is small relative to the ground state energy, $ E_A^{(0)} + E_{B_i}^{(0)} , $ of adjacent, but unperturbed $ \mathcal{A} $ and $ \mathcal{B}_i , $  the London collision model \cite{london, hirschfelder} is linear and can be readily modified by superposition to account for n-body interactions in which $ \mathcal{A} $ simultaneously experiences dispersive interactions with $ n $ neighboring molecules. Generalizing Hirschfelder \cite{hirschfelder} by assuming a perturbed wave function that is the product  of the $ n $ unperturbed, isolated wave functions for $ n$  colliding molecules, it is readily shown that the approximate, second-order, dispersive potential has the form, 
\begin{displaymath}\label{phimultiatom}
\phi^{(n)} \left( \mathbf{r_A} , \mathbf{r_2} , ..., \mathbf{r_{n-1}} \right) = -\left( 3/2 \right) E_I \alpha^2 \left[ \mathbf{r_{A1}}^{-6} + \mathbf{r_{A2}}^{-6} + ... + \mathbf{r_{Am}}^{-6} \right] 
 \end{displaymath}
where $ E_I , $ an empirical constant, is on the order of the ionization energy, $ \alpha $ is the polarizability, $ \mathbf{r_{Ai}} $ is the internuclear distance between molecules $ \mathcal{A} $ and $ \mathcal{B}_i , $ and $ m=n-1 .$ 

A 'simultaneous n-body collision' takes place when the internuclear distances between $ \mathcal{A} $ and $n-1$ immediately adjacent molecules are all approximately equal to the minimum of these distances, $ \mathbf{r_{A1}} \approx \mathbf{r_{A2}} \approx ... \approx \mathbf{r_{Ai_{min}}}. $ Writing $ \mathbf{r_{Aj}} = \mathbf{r_{Ai_{min}}} + \Delta \mathbf{r_{Aj}} ,$ forming the ratio $ \mathbf{r_{Aj}} / \mathbf{r_{Ai_{min}}} ,$ and Taylor expanding $ \mathbf{r_{Aj}} , $ we see that for an n-body collision to occur - corresponding to $ n $ non-negligible contributions to $ \phi^{(n)} \left( \mathbf{r_A} , \mathbf{r_2} , ..., \mathbf{r_{n-1}} \right) $- all $ n-1 $ molecules must remain within approximately $ 16 \% $ of the minimum separation, $ \mathbf{r_{Ai_{min}}}. $ Thus, while three-body collisions certainly take place, for example, due to this restrictive condition, pair-wise collisions dominate. Predicted dynamic viscosities above, which assume dominant pairwise collisions, are consistent with this simple argument. 

\begin{center}
\begin{table*}[t]
\begin{adjustwidth}{-2.0in}{0in}
\begin{tabular}{|c|c|c|c|c|}
\hline
\hline
\multicolumn{5} {|c|} {}  \\
\multicolumn{5} {|c|} {\Large{Single molecule dynamics models}} \\
\multicolumn{5} {|c|} {} \\
\hline
\hline
 \large{Time scale} & \large{Dominant physics} & \large{External} & \large{Friction force} & \large{Random force} \\
\large{[Collective dynamics]}  & \large{[Example]} & \large{force} & \large{(deterministic)} &    \\
\hline
$ \tau_d \leq t \leq \tau_F $ & Phonons; low dissipation &  &  &   \\
 \small{[Elastic]} & [Continuum flow]  & N/A &
$ -  3 \pi \sigma_m \mu \mathbf{v} \left( t \right) \rightarrow 0 $ &  $ -M \sum_{i=1} \omega_i^2 \int_{0}^t \tilde{\mathbf{v}}_i \left( t , \tau \right) d \tau $  \\
\hline
$ \tau_d \leq t \leq \tau_F $ & Phonons; dissipation &  &  &   \\
 \small{[Viscoelastic]} & [Particle scattering] & $ \mathbf{F} \left( t \right) $ &
$ - \int_0^t \kappa \left( t - t' \right) \mathbf{v} \left( t' \right) dt' $ & $ - M \sum_{i=1} \omega_i^2 \int_{0}^t \tilde{\mathbf{v}}_i \left( t , \tau \right) d \tau $ \\
\hline
$ \tau_F \leq t \leq \tau_c $ & Phonons; low dissipation &  &  &  \\
 \small{[Transition]} & [Continuum flow]  & N/A &
$ -  3 \pi \sigma_m \mu \mathbf{v} \left( t \right)  \rightarrow 0 $ & see remark d) \\
\hline
$ \tau_F \leq t \leq \tau_c $ & Phonons; dissipation &  &  &  \\
 \small{[Viscoelastic]} & [Particle scattering]  & $ \mathbf{F} \left( t \right) $ &
$ - \int_{0}^t \kappa \left( t - t' \right) \mathbf{v} \left( t' \right) dt' $ & see remark d) \\
\hline
$ t \geq \tau_c $ & Brownian force; dissipation &  &  &  \\
 \small{[Fluid]} & [Continuum flow] & N/A &
$ - 3 \pi \sigma_m \mu \mathbf{v} \left( t \right) $ & $ \dot{\eta} \left(t \right) $ \\
\hline
$ t \geq \tau_c $ & Brownian force; dissipation &  &  &  \\
 \small{[Viscoelastic]} & [Particle scattering] & $ \mathbf{F} \left( t \right) $ &
$ - \int_0^t \kappa \left( t - t' \right) \mathbf{v} \left( t' \right) dt' $ &  $ \dot{\eta} \left(t \right) $ \\
\hline
\end{tabular}
\caption{\label{tab:table1}Situation- and time-scale-dependent force terms can be inserted into the generic Langevin equation, (\ref{langevineqngeneric}). Explanatory notes regarding each force term are given above as points a) through i).}
\end{adjustwidth}
\end{table*}
\end{center}

\section*{Conclusions}
Unraveling the dynamics of individual atoms and  small molecules in liquids represents a centuries-old physics problem.  While neutron and light-scattering experiments, as well as molecular dynamics simulations, instantaneous normal mode analyses, and molecular hydrodynamics expose and explain single-particle-scale and collective liquid-state dynamics, the description is largely couched in terms of dynamical correlation functions.  In an attempt to expose the essential dynamical elements that  determine single molecule motion, at least in nonpolar liquids, this paper presents physical arguments that suggest: i) intermolecular dispersion forces and temperature-dependent electron screening determine viscosity, i.e., temperature-dependent intermolecular friction forces, and ii) a narrow band of phonons, lying near the liquid-solid (Frenkel) transition frequency, drives the random molecular jumps constituting self-diffusion.

In mechanistic terms, we present preliminary evidence that, in simple liquids, both viscosity and single molecule viscous drag emerge due to small, collision-induced \textit{tangential} distortions of individual electron clouds. By contrast, self-diffusional, single-molecule hops are produced by collision-induced \textit{compression} of interacting molecular clouds; the latter mechanism pushes interacting molecular pairs into short-lived repulsive energy states. 

We are hopeful that the preliminary picture of single molecule, liquid state dynamics proposed here promotes further progress in understanding this complex problem.


\bibliography{library1}

\bibliographystyle{abbrv}

\section*{Acknowledgments}
This material is based upon research supported by, or in part by, a grant from  the U. S. Office of Naval Research under award number N00014-18-1-2754.

\end{document}